\documentclass[aps,pra,preprint,groupedaddress]{revtex4}

\usepackage{amssymb,amsmath,epsfig,natbib}
\usepackage{hyperref}
\usepackage{cleveref}
\usepackage{subfigure}

\newcommand{\RR}{{\mathbb{R}}}

\newcommand{\ZZ}{{\mathbb{Z}}}
\newcommand{\pa}{\partial}
\newcommand{\dd}{{\rm d}}
\newcommand{\Tr}{\mathrm{Tr}}
\newcommand{\SU}{\mbox{SU}}

\newcommand{\bphi}{\mathbf{\phi}}

\begin{document}

\title{Baby Skyrme models without a potential term}

\author{Jennifer Ashcroft}
\email[]{J.E.Ashcroft@kent.ac.uk}

\affiliation{School of Mathematics, Statistics and Actuarial Science, University of Kent, Canterbury, CT2 7NF, UK}

\author{Mareike Haberichter}
\email[]{M.Haberichter@kent.ac.uk}
\affiliation{School of Mathematics, Statistics and Actuarial Science, University of Kent, Canterbury, CT2 7NF, UK}

\author{Steffen Krusch}
\email[]{S.Krusch@kent.ac.uk}
\affiliation{School of Mathematics, Statistics and Actuarial Science, University of Kent, Canterbury, CT2 7NF, UK}

\date{\today \\ \vspace*{3cm}}

%%%%%%%%%%%%%%%%%%%%%%%%%%%%%%%%%%%%%%%%%%%%%%%%%%%%%%%%%%%%%%%%%%%%%%%%%%%%%%%%%%%%%%%%
\begin{abstract}
We develop a one-parameter family of static baby Skyrme models that do not require a potential term to admit topological solitons. This is a novel property as the standard baby Skyrme model must contain a potential term in order to have stable soliton solutions, though the Skyrme model does not require this. Our new models satisfy an energy bound that is linear in terms of the topological charge and can be saturated in an extreme limit. They also satisfy a virial theorem that is shared by the Skyrme model. We calculate the solitons of our new models numerically and observe that their form depends significantly on the choice of parameter. In one extreme, we find compactons while at the other there is a scale invariant model in which solitons can be obtained exactly as solutions to a Bogomolny equation. We provide an initial investigation into these solitons and compare them with the baby Skyrmions of other models. 
\end{abstract}

%%%%%%%%%%%%%%%%%%%%%%%%%%%%%%%%%%%%%%%%%%%%%%%%%%%%%%%%%%%%%%%%%%%%%%%%%%%%%%%%%%%%%%%%

\maketitle

%%%%%%%%%%%%%%%%%%%%%%%%%%%%%%%%%%%%%%%%%%%%%%%%%%%%%%%%%%%%%%%%%%%%%%%%%%%%%%%%%%%%%%%%
%%%%%%%%%%%%%%%%%%%%%%%%%%%%%%%%%%%%%%%%%%%%%%%%%%%%%%%%%%%%%%%%%%%%%%%%%%%%%%%%%%%%%%%%

\section{Introduction}\label{sec:intro}

The baby Skyrme model \cite{Piette:1994ug,Piette:1994mh} is a nonlinear field theory admitting topological solitons known as baby Skyrmions. It is often studied as a (2+1)-dimensional analogue of the Skyrme model \cite{Skyrme:1961vq} for nuclear physics though is itself an interesting physical model with applications in condensed matter physics \cite{Sondhi:1993,Walet:2001zz,Yu:2010}. In the (3+1)-dimensional Skyrme model, the topological solitons are called Skyrmions and can be used to model atomic nuclei with their topological charge, an integer $B$, giving the baryon number. As a lower-dimensional version of this model, the baby Skyrme model has been used to investigate a variety of difficult problems in the Skyrme theory including Skyrmion scattering \cite{Peyrard:1990hc, Piette:1994mh, Gisiger:1998tv, Foster:2014vca} and the effect of isorotation on Skyrmion solutions \cite{Battye:2013tka, Halavanau:2013vsa, Battye:2014qva}. 

A key difference in the models arises when we consider their necessary components. The baby Skyrme model is an $O(3)$-sigma model extended by the addition of a term quartic in derivatives called the Skyrme term and a symmetry breaking potential term. The combination of Skyrme term and potential gives a scale to the model and enables it to evade Derrick's theorem \cite{Derrick:1964} for scalar field theories in two space dimensions. By contrast, in the full Skyrme model the combination of the Skyrme and sigma terms is sufficient to evade Derrick's theorem. 
This provides one motivation for our paper --- we wish to design a static baby Skyrme model that does not require a potential term to have topological solitons.
One approach is to apply a noncommutative deformation to the baby Skyrme model instead of including a potential term \cite{Ioannidou:2009re,Domrin:2013ega}. We apply a very different method.

Before discussing our approach in detail, we briefly review 
different variants of the baby Skyrme model. One way to create new models has been through the use of different potentials 
 \cite{Eslami:2000tj,Weidig:1998ii,Jaykka:2010bq,Hen:2007in}, and it has been found that the choice of potential has a dramatic effect on the solitons of the model. In particular, the appearance and structure of multisolitons depends strongly on the potential term. For some potentials, higher-charge solitons form chains \cite{Foster:2009vk}, for some rings \cite{Hen:2007in} and for others \cite{Leese:1989gi,Sutcliffe:1991aua} stable multisolitons may not exist at all. Models have also been designed in which the $O(3)$ symmetry is broken to the dihedral group $D_N$, and here multi-Skyrmions have been observed with crystalline or broken structures \cite{Ward:2003xv,Jaykka:2011ic,Winyard:2013ada}.

In addition to choosing a different potential term, it is possible to develop new baby Skyrme models by removing the sigma term. Models consisting of only the Skyrme term and a potential are sometimes called restricted or BPS baby Skyrme models \cite{Gisiger:1996vb,Adam:2010jr,Adam:2012hh}. Deformations of BPS models \cite{Bolognesi:2014ova} have also been investigated, for which a physical motivation is found in the (3+1)-dimensional Skyrme theory. One significant problem in applying the Skyrme model to nuclear physics is that the binding energies of Skyrmions are considerably larger than the experimental values. The BPS Skyrme model \cite{Adam:2010fg,Adam:2014loa,Adam:2012za} consists only of a sextic term and a potential term, and has been developed, along with its generalizations, as an attempt to obtain more realistic binding energies. Different approaches to obtain Skyrme models with low binding energies are by coupling vector mesons to the $O(3)$-sigma model term and removing the Skyrme term \cite{Sutcliffe:2008sk,Foster:2009rw} or by studying Skyrme solitons on curved backgrounds \cite{Manton:1986pz,Canfora:2014aia}.

Another attempt to address the problem of obtaining realistic classical binding energies in the Skyrme theory has been to create new Skyrme models through a novel choice of potential term \cite{Harland:2013rxa,Gillard:2015eia}. A family of models is obtained by a one-parameter family of potential functions interpolating between the standard Skyrme model and a model in which a topological energy bound can be saturated for $|B|=1$. An equivalent idea has been explored in the baby Skyrme model to obtain so-called ``aloof" baby Skyrmions \cite{Salmi:2014hsa}. We have also been motivated by recent interest in topological energy bounds \cite{Adam:2013tga,Harland:2013rxa}. When designing our models, we require that they satisfy a particular topological energy bound. We find that this has several useful consequences for our models.

Our approach to designing new baby Skyrme models is entirely different from those outlined above. As we wish to design models which do not require a potential term to have topological solitons, we remove the potential and raise the sigma and Skyrme terms to some powers. Considering Derrick's scaling argument and requiring that our models satisfy a topological energy bound results in a one-parameter family of baby Skyrme models. We find that the required powers of the Skyrme and sigma terms are fractional. This draws a natural comparison between our models and the Nicole \cite{Nicole:1978nu} and AFZ \cite{Aratyn:1999cm,Aratyn:1999cf} models, which were investigated numerically in Refs. \cite{Gillard:2010xi,Gillard:2010qda} along with a set of conformally invariant Skyrme-Faddeev models obtained by taking linear combinations of the two.

The outline of the paper is as follows. We open in Sec. \ref{sec:bsky} with a brief overview of the static baby Skyrme model, focusing on its well-known energy bound and the application of the Derrick scaling argument to this theory. In Sec. \ref{sec:main} we present our new models, beginning with a general form for the static energy and then illustrating how the application of Derrick's theorem and the requirement that our solitons satisfy a topological energy bound reduces the number of parameters to one.  

In the remainder of the paper we investigate the solitons of our one-parameter family of models. In Sec. \ref{sec:numerics} we present our numerical results, first considering axially symmetric solutions and then progressing to simulations of the full field theory. We discuss the effect of the model parameter on the solitons and compare them to those found in existing models. We end by summarizing our results and reflecting upon open questions and opportunities for further investigation.

%%%%%%%%%%%%%%%%%%%%%%%%%%%%%%%%%%%%%%%%%%%%%%%%%%%%%%%%%%%%%%%%%%%%%%%%%%%%%%%%%%%%%

\section{The Baby Skyrme model}\label{sec:bsky}
The static energy functional of the baby Skyrme model is given by
\begin{equation}
\label{bs-energy}
E_{BS} = \int_{\RR^{2}} \bigg(\frac{1}{2}\pa_i \bphi \cdot \pa_i \bphi + \frac{1}{4}|\pa_i \bphi \times \pa_j \bphi |^2+ V(\bphi) \bigg)~\dd^2 x,
\end{equation}
where the field $\bphi:\RR^2 \to S^2$ is a three-component vector $\bphi = (\phi_1,\phi_2,\phi_3)$ of unit length. The first term in \eqref{bs-energy} is the $O(3)$-sigma model term and is extended by the addition of a term quartic in derivatives, called the Skyrme term, and a potential term $V(\bphi)$ to allow the existence of stable topological soliton solutions.

To ensure that solutions have finite energy, the boundary condition 
\begin{align}\label{bc}
\lim_{|x|\rightarrow\infty} ~\bphi = (0,0,1),
\end{align}
is imposed, assuming that $(0,0,1)$ is a minimum of the potential $V$. This enables a one-point compactification $\RR^2\cup\{\infty\} \cong S^2$, and thus we can consider $\bphi$ as a map $\bphi:S^2\rightarrow S^2$. We can label the maps $\bphi$ by an integer $B\in\pi_2(S^2)=\ZZ$, called the topological charge. This is the winding number of the map, given by
\begin{align}\label{charge} 
B=-\frac{1}{4\pi} \int_{\RR^2} \bphi\cdot(\pa_1\bphi\times\pa_2\bphi)~ \dd^2x,
\end{align}
and is sometimes called the baryon number for comparison with the Skyrme model. The topological solitons of this theory are field configurations which minimize the energy \eqref{bs-energy} in a given topological sector $B$. They are called baby Skyrmions.

A lower bound on the energy of a solution with charge $B$ in the baby Skyrme model is given by
\begin{align}\label{Bsky_bound}
E_{BS}\geq4\pi|B|.
\end{align}
This is a bound on the sigma term alone, obtained by a completing the square argument, and is never saturated by baby Skyrmions. 

When deriving energy bounds, it will be convenient for us to rewrite the static energy \eqref{bs-energy} using its geometrical interpretation \cite{Manton:1987xt}. Define the symmetric, positive definite $2\times2$ matrix $D$ by 
\begin{align}\label{D_matrix}
D_{ij}=\pa_i\bphi\cdot\pa_j\bphi,
\end{align}
and let $\lambda_i^2$ denote the eigenvalues of the strain tensor $D$, where $i=1,2$. Then we can express the baby Skyrme energy functional \eqref{bs-energy} in terms of the non-negative eigenvalues of $D$ as
\begin{align}\label{bs-eigs}
E_{BS} &= \int_{\RR^{2}} \bigg(\frac{1}{2}(\lambda_1^2+\lambda_2^2)  + \frac{1}{2}\lambda_1^2\lambda_2^2+ V(\bphi) \bigg)\dd^2 x,
\end{align}
and the topological charge can be expressed as
\begin{align}\label{charge-eigs}
B=-\frac{1}{4\pi} \int_{\RR^2} \lambda_1\lambda_2~ \dd^2x.
\end{align}
Using the energy \eqref{bs-eigs}, we can obtain the well-known lower energy bound \eqref{Bsky_bound} on the sigma term by completing the square as
\begin{align}\label{bs_bound_eigs}
\frac{1}{2}\int_{\RR^{2}} (\lambda_1^2+\lambda_2^2) ~\dd^2x & = \frac{1}{2}\Big(\int_{\RR^{2}} (\lambda_1\mp\lambda_2)^2 ~\dd^2x\pm2\int_{\RR^{2}} \lambda_1\lambda_2 ~\dd^2x \Big), \nonumber \\
 & \geq \Big|\int_{\RR^{2}} \lambda_1\lambda_2 ~\dd^2x\Big| = 4\pi|B|.
\end{align}
The approach given above for deriving topological energy bounds is similar to those given in recent papers on the subject \cite{Adam:2013tga,Harland:2013rxa}. We will apply this method again in Sec. \ref{sec:bounds} when we derive energy bounds for our new baby Skyrme models. While \eqref{Bsky_bound} is a well-known topological energy bound for the baby Skyrme model, recently tighter bounds have been obtained by also taking into account energy contributions from the Skyrme term and the potential term \cite{Adam:2010jr,Adam:2013tga}. 

The inclusion of  a potential term in the baby Skyrme model is important as it allows the model to evade Derrick's theorem \cite{Derrick:1964} and thus have topological soliton solutions. This theorem rules out the existence of topological solitons in flat space scalar field theories by the requirement that a stationary point of the energy must also be stationary against rescaling. Therefore, if the energy of the theory after applying the spatial rescaling $x\mapsto\mu x$, which we denote by $e(\mu)$, has no stationary point, then there can be no static finite energy solutions except the vacuum.

We apply this argument to the baby Skyrme model. Under the rescaling $x \mapsto \mu x$, the static energy \eqref{bs-energy} becomes
\begin{align}
e_{BS}(\mu)&= E_2+\mu^2 E_4 +\mu^{-2} E_0,
\end{align}
where we use $E_2,~E_4$ and $E_0$ to denote the sigma term, Skyrme term and potential term, respectively. As a result of Derrick's theorem, we observe that the combination of a potential term and the Skyrme term allows the existence of topological solitons.

We can also derive a virial theorem satisfied by the baby Skyrme model by taking $\frac{de_{BS}}{d\mu}|_{\mu=1}$ and setting this to zero, to find
\begin{align}\label{orig_virial}
  E_4 = E_0.
\end{align}

Contrast this with the results of applying the scaling argument to the Skyrme model, which has static energy
\begin{align}
\label{Skyrmion-energy}
E_S &= \int_{\RR^{3}} \bigg(-\frac{1}{2}\Tr(R_iR_i) - \frac{1}{16}\Tr([R_i,R_j][R_i,R_j]) + m^2\Tr(1-U) \bigg)~\dd^3 x,
\end{align}
where $m$ is related to the pion mass, the pion fields are written as $U:\mathbb{R}^3 \to \SU(2)$ and $R_i = (\pa_i U)U^\dag$.

In this case, applying the rescaling produces
\begin{align}
e_S(\mu)&= \frac{1}{\mu}E_2+\mu E_4 +\frac{1}{\mu^3} E_0,
\end{align}
so the potential term is unnecessary to evade Derrick's theorem. If we consider the static energy of the Skyrme model with no potential term, we can further obtain the virial theorem \begin{align}\label{Skyrme_virial}
E_2=E_4.
\end{align}

We have seen that the potential term is a necessary component of the baby Skyrme model if there are to exist topological soliton solutions. However, the same is not true of the Skyrme model, in which Skyrmions can exist without the presence of a potential term. This difference between the two theories motivates us to investigate the design of baby Skyrme models that do not include a potential term but still have soliton solutions.

A variety of different functions $V(\bphi)$ have been investigated as the potential term in the baby Skyrme model \cite{Eslami:2000tj,Weidig:1998ii,Jaykka:2010bq,Hen:2007in}.  Particular examples are
\begin{align}\label{potentials}
	V(\bphi)= \left\{
		\begin{array}{lll}
			V_1=m^2(1-\phi_3) & \textnormal{(``old" potential)} \\
			V_2=m^2(1-\phi_3^2) & \textnormal{(``new" potential)} \\
			V_3=m^2(1-\phi_3)^4 & \textnormal{(holomorphic potential).}
		\end{array}
	\right.
\end{align}
 The choice of potential has a strong effect on the appearance and structure of multisoliton solutions. For example, in the old baby Skyrme model $V_1$, higher-charge baby Skyrmions form chains \cite{Foster:2009vk}, in the new baby Skyrme model $V_2$ rings are minima, and in the holomorphic model $V_3$ no multisolitons exist. Recent work \cite{Salmi:2014hsa} has explored combining the old potential $V_1$ with the holomorphic potential $V_3$ to obtain weakly bound multisolitons. We are interested in what the structure of baby Skyrmions would be without a potential to govern them.

%%%%%%%%%%%%%%%%%%%%%%%%%%%%%%%%%%%%%%%%%%%%%%%%%%%%%%%%%%%%%%%%%%%%%%%%%%%%%%%%%%%%%%%%%

\section{Baby Skyrme models without a potential}\label{sec:main}
We propose a range of new baby Skyrme models that do not require a potential term to evade Derrick's theorem. To achieve this, we raise the sigma and Skyrme terms to the power $\alpha$ and $\beta$, respectively, and determine the range of acceptable values for these powers to ensure  stability with respect to rescaling. As a starting point for the new static energy, take 
\begin{align}
\label{energy-alphabeta}
E &= \int_{\RR^{2}} \Big(c_1 (\pa_i \bphi \cdot \pa_i \bphi)^{\alpha} + c_2 (|\pa_i \bphi \times \pa_j \bphi|^2)^{\beta} \Big)~\dd^2 x,
\end{align}
where $c_1,~c_2$ are positive real coupling constants, and $\alpha,~\beta$ are real constants. 

%--------------------------------------------------------------------------------------

\subsection{Derrick's scaling argument}\label{sec:Derrick}
To determine suitable values for $\alpha$ and $\beta$, we apply the rescaling  $x \mapsto \mu x$ to the static energy \eqref{energy-alphabeta} and consider the results of Derrick's theorem. This leads to the energy
\begin{align}
e(\mu)&= \mu^{2\alpha-2} E_2+\mu^{4\beta-2} E_4.
\end{align}
There are three cases in which our model can evade Derrick's theorem:
\begin{equation}\label{alphabeta-constraint}
\begin{split}
\textnormal{(i)}~~~ & \alpha<1~~\textnormal{and}~~\beta>0.5, \\
\textnormal{(ii)}~~~ & \alpha>1~~\textnormal{and}~~\beta<0.5, \\
\textnormal{(iii)}~~~ & \alpha=1~~\textnormal{and}~~\beta=0.5,
\end{split}
\end{equation}
with case (iii) providing a scale invariant model. We only consider cases (i) and (iii) because solutions in the models of case (ii) would either be compact or not have finite energy; see Appendix \ref{app:near_f0} for a detailed discussion.

There is also a virial theorem satisfied by our models. Take $\frac{de}{d\mu}|_{\mu=1}$ and set this equal to zero. For case (iii) this is automatically satisfied; otherwise, we have
\begin{align}\label{virial_calc}
(2\alpha-2) E_2+(4\beta-2) E_4 = 0.
\end{align}
The resulting virial theorem is
\begin{align}\label{virial}
 E_2=\frac{1-2\beta}{\alpha-1}~ E_4 .
\end{align}

Recall that the Skyrme model without a potential term satisfies the virial theorem $E_2~=~E_4$. Our models also satisfy this virial theorem when 
\begin{align}\label{virial_constraint}
\beta=1-\frac{\alpha}{2}.
\end{align}
This selection of models includes one in which the static energy \eqref{energy-alphabeta} produces the same function $e_S(\mu)$ under rescaling as that for the Skyrme model \eqref{Skyrmion-energy} without a potential term. In this case the parameters are $\alpha=0.5$ and $\beta=0.75$.

%---------------------------------------------------------------------------------------

\subsection{Energy bounds}\label{sec:bounds}
We have seen in Sec. \ref{sec:bsky} that the baby Skyrme model \eqref{bs-energy} satisfies a linear bound \eqref{Bsky_bound} on the energy of its solutions in terms of the number of solitons. This is a useful property shared by many soliton models \cite{manton2004topological}. Therefore, we require our new baby Skyrme models to satisfy such a lower bound on the energy. In the following, we use this condition to fix the parameter $\beta$ in \eqref{energy-alphabeta} and further restrict the family of models that we consider.

Defining the matrix $D$ by $D_{ij}=\pa_i\bphi\cdot\pa_j\bphi$ as before, rewrite the energy \eqref{energy-alphabeta} as
\begin{equation}\label{eneval}
E = \int_{\RR^{2}} \Big(c_1 (\lambda_1^2+\lambda_2^2)^{\alpha} + c_2 (2\lambda_1^2\lambda_2^2)^{\beta} \Big)~\dd^2 x,
\end{equation}
where $\lambda_1^2,~\lambda_2^2$ denote the eigenvalues of $D$. 

To obtain a lower bound on the energy, we use the following special case of the inequality of the arithmetic and geometric means: for $a$, $b$ non-negative, 
\begin{equation}\label{ineq-halfsqrt}
\frac{a+b}{2}\geq\sqrt{ab},
\end{equation}
with equality if and only if $a=b$.

We obtain a lower bound on the energy by first applying inequality \eqref{ineq-halfsqrt} twice to find
\begin{equation}\label{genbound}
\begin{split}
E & = 2\int_{\RR^{2}} \Big(\frac{1}{2} c_1 (\lambda_1^2+\lambda_2^2)^{\alpha} + \frac{1}{2} c_2 (2\lambda_1^2\lambda_2^2)^{\beta}\Big)\dd^2 x, \\
& \geq 2\int_{\mathbb{R}^2}\sqrt{c_1}(\lambda_1^2+\lambda_2^2)^{\frac{\alpha}{2}}\sqrt{c_2}(2\lambda_1^2\lambda_2^2)^{\frac{\beta}{2}}\dd^2x, \\
& \geq 2^{1+\frac{\beta}{2}}\sqrt{c_1c_2}\int_{\mathbb{R}^2}2^{\frac{\alpha}{2}}|\lambda_1\lambda_2|^{\frac{\alpha}{2}}|\lambda_1\lambda_2|^{\beta} \dd^2x, \\
& = 2^{1+\frac{\alpha+\beta}{2}}\sqrt{c_1c_2}~\int_{\mathbb{R}^2}|\lambda_1\lambda_2|^{\frac{\alpha+2\beta}{2}}~\dd^2x.
\end{split}
\end{equation}
Then to ensure that this energy bound is linear in terms of the topological charge $B$, the required value of $\beta$ is
\begin{equation}\label{boundreq2}
\beta=1-\frac{\alpha}{2}.
\end{equation}
The resulting topological energy bound is
\begin{equation}\label{generalbound}
E \geq 2^{\frac{3}{2}+\frac{\alpha}{4}}\sqrt{c_1c_2}~4\pi|B|.
\end{equation}
Note that this choice of $\beta$ was also found in Sec. \ref{sec:Derrick} by requiring that the virial theorem \eqref{virial} be simply $E_2=E_4$: the virial theorem of the Skyrme model without a potential term~\footnote{Note that we can generalise our topological energy bound by splitting up the second term in the energy \eqref{eneval} into $M$ components and using the inequality of the arithmetic and geometric means for $M$$+$$1$ components. This results in the condition $\beta=(M+1-\alpha)/2M$. Our bound is reproduced when setting $M=1$, in which case the constraint on $\beta$ is the same as that required by the virial theorem $E_2=E_4$. (Private communication with  A.~Wereszczynski)}. 

%% alpha=1 bound

In case (iii), there is an alternative energy bound derived by a standard completing the square argument. The static energy for this model is given in terms of the eigenvalues $\lambda_1^2,~\lambda_2^2$ as
\begin{align}\label{a1en_eigs}
E = \int_{\RR^{2}} \Big(c_1 (\lambda_1^2+\lambda_2^2) + c_2 \sqrt{2\lambda_1^2\lambda_2^2} \Big)\dd^2 x.
\end{align}
By completing the square, we find
\begin{align}\label{a1en_eigs_working}
E & = \int_{\RR^{2}} \Big( c_1 (|\lambda_1|-|\lambda_2|)^2 +2c_1|\lambda_1\lambda_2| + \sqrt{2}c_2 |\lambda_1\lambda_2| \Big) \dd^2 x,  \nonumber \\
& = \int_{\RR^{2}} c_1 (|\lambda_1|-|\lambda_2|)^2 \dd^2 x + (2c_1+\sqrt{2}c_2)\int_{\RR^{2}} |\lambda_1\lambda_2|~ \dd^2 x ,  \nonumber \\
& \geq  4\pi(2c_1+\sqrt{2}c_2)|B|.
\end{align}
So an alternative bound in case (iii) is given by 
\begin{align}\label{a1_bound}
E\geq 4\pi(2c_1+\sqrt{2}c_2)|B|.
\end{align}
This bound is saturated for solutions of the Bogomolny equation, 
\begin{align}\label{Bog_eq_lambda}
|\lambda_1|=|\lambda_2|,
\end{align}
which leads to the following system of equations:
\begin{subequations}\label{bog_syst}
\begin{eqnarray}
\pa_1\bphi\cdot\pa_1\bphi &  =  &  \pa_2\bphi\cdot\pa_2\bphi, \\ 
\pa_1\bphi\cdot\pa_2\bphi &  =  &  0.  
\end{eqnarray}
\end{subequations}
Let $z$ denote the complex coordinate $z=x+iy$ in the spatial plane and $R$ the Riemann sphere coordinate on the target $S^2$. We can write solutions of Eqs. \eqref{bog_syst} in each topological sector $B$ in terms of rational maps $R(z)$ as
\begin{align}\label{symm_a1_rat}
\bphi=\frac{1}{1+|R|^2}\Big(R+\bar{R}, ~-i(R-\bar{R}),~ |R|^2-1 \Big).
\end{align}
Here $R(z)=p(z)/q(z)$ is a ratio of two polynomials $p(z)$ and $q(z)$ with no common factors, and $B=\max\{\deg(p),\deg(q)\}$. To satisfy the boundary condition $\lim_{|x|\rightarrow\infty} ~\bphi = (0,0,1)$ at infinity, we require $R(\infty)=\infty$. 
One important case is the axially symmetric rational map
\begin{align}\label{R(z)}
R(z)=z^B.
\end{align}
 Thus, we can find exact solutions for baby Skyrmions of any charge $B$ in this model.

%% fixing the constants

It still remains to set the values of the coupling constants $c_1$ and $c_2$. In this paper, we choose
\begin{align}\label{c1c2}
c_1=2^{-\frac{3+\alpha}{2}} \quad \textnormal{and} \quad
  c_2=2^{\frac{\alpha}{2}}c_1,
 \end{align}
 and so obtain the final form of the static energy for our models as
\begin{align}\label{final_en}
E=2^{-\frac{3+\alpha}{2}}\int_{\RR^{2}} \Big( (\pa_i \bphi \cdot \pa_i \bphi)^\alpha + 2 (|\pa_1 \bphi \times \pa_2 \bphi|^2)^{1-\frac{\alpha}{2}} \Big)\dd^2 x.
\end{align}
This choice of coupling constants has three useful consequences. First, the choice of $c_2$ ensures that the bounds \eqref{generalbound} and \eqref{a1_bound} in case (iii) coincide. It also causes the virial theorem $E_2=E_4$ to be satisfied in case (iii), as we now have $c_1=\frac{1}{4}$ and $c_2=\frac{1}{2\sqrt{2}}$ and thus,
\begin{align}
E_2=2c_1\cdot4\pi|B|=2\pi|B|=\sqrt{2}c_2\cdot 4\pi|B|=E_4.
\end{align}
Due to the scale invariance of this model, it is not necessary that the virial theorem be satisfied in this case. It is only due to the choice of constants \eqref{c1c2} that the virial theorem \eqref{Skyrme_virial} holds here.

Finally, for any choice of $\alpha$, this combination of $c_1$ and $c_2$ sets the topological energy bound \eqref{generalbound} to be
\begin{align}\label{4pi_bound}
E\geq 4\pi |B|,
\end{align}
with the bound saturated when $\alpha=1$. This is the well-known energy bound on the sigma term of the baby Skyrme model.

%%%%%%%%%%%%%%%%%%%%%%%%%%%%%%%%%%%%%%%%%%%%%%%%%%%%%%%%%%%%%%%%%%%%%%%%%%%%%%%%%%%%%%%%%

\section{Numerical Results}\label{sec:numerics}

In this section, we calculate axially symmetric baby Skyrmion solutions for parameter $\alpha\in~[0.5,1]$ and with topological charges $B=1-3$,~10. We minimize the energy functional (\ref{final_en}) for rotationally symmetric Skyrme configurations using two very different numerical approaches: 1D gradient flow and Newton's method for nonlinear systems. Finally, we perfom 2D energy minimization simulations for a selection of our models and verify that the minimal energy solutions agree with those found when imposing axial symmetry. Skyrmion chain solutions are found to be of higher energy. 

Note that the baby Skyrmion solutions for $\alpha=0.5$ are discussed in a separate subsection as the solitons become \emph{compactons}. These are solitons with compact support, taking vacuum values everywhere outside some finite region of space. Compact solitons have been studied before in the Skyrme-Faddeev model in the infinite mass limit \cite{Foster:2010zb}, and in massive baby Skyrme models \cite{Adam:2009px,Adam:2010jr,Speight:2010sy}.  Compactons are numerically challenging and require a careful adjustment of our numerical methods.

%----------------------------------------------------------------------------------------
\subsection{Axial baby Skyrme solutions}\label{sec:axial}
To find axially symmetric soliton solutions of the equations of motion, we use the ansatz
\begin{align}
\label{ansatz}
\bphi=(\sin f \cos (B\theta),~ \sin f \sin (B\theta),~ \cos f),
\end{align}
where $r$, $\theta$ are the usual polar coordinates, $f=f(r)$ is a radial profile function, and $B$ is the topological charge of the configuration. Substituting \eqref{ansatz} into \eqref{final_en} yields the energy
\begin{align}
\label{ansatz-energy}
E =  2^{-\frac{3+\alpha}{2}} \cdot 2\pi \int_0^\infty \Big( \left(f'^2+B^2\frac{\sin^2 f}{r^2} \right)^\alpha+2\left(f'^2B^2\frac{\sin^2 f}{r^2}\right)^{1-\frac{\alpha}{2}} \Big) ~r \dd r,
\end{align}
 which depends only on the radial coordinate $r$. Here prime denotes differentiation with respect to the radial coordinate $r$. By the principle of symmetric criticality, solutions of the Euler-Lagrange equation for the simplified energy \eqref{ansatz-energy} will also solve the equations of motion for the original energy \eqref{final_en}. In the following, we minimize the energy \eqref{ansatz-energy} by solving the Euler-Lagrange equation \eqref{symm-eqm-alphabeta} in Appendix \ref{app:near_f0} subject to the boundary conditions $f(0)=\pi$ and $f(\infty)=0$ in two ways: through the use of a 1D flow method and also using Newton's method for nonlinear systems \cite{NumRec:2007} with grid spacing $\Delta r=10^{-4}$ over the interval $0\leq  r \leq20$.  For both methods, we monitor the topological charge and check the virial theorem~\eqref{virial} at each iteration step.

\begin{figure}
   \subfigure[]{\label{fig:energy_comparison_newcs} \includegraphics[height=5.5cm]{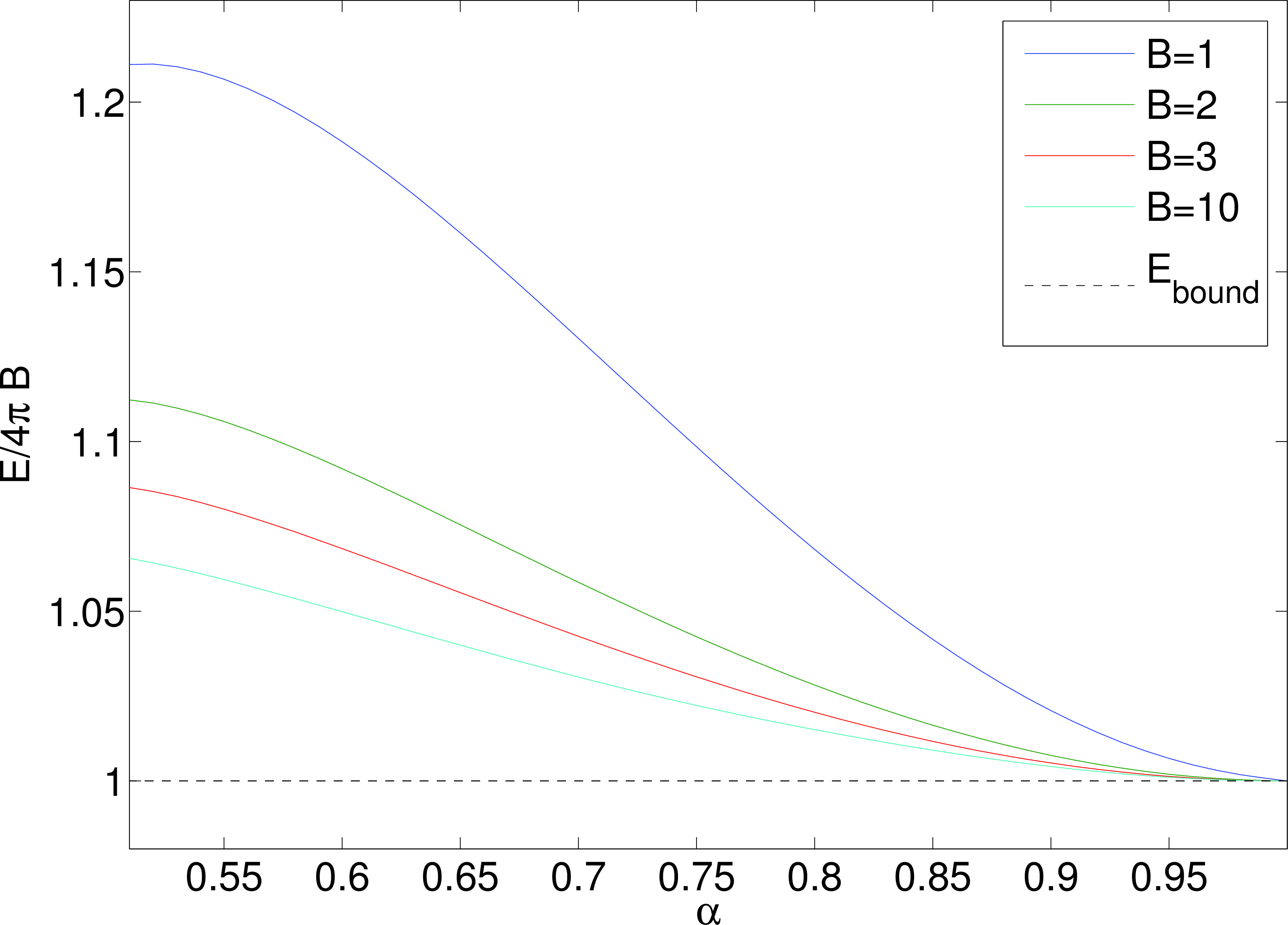} }
          \subfigure[]{\label{fig:binding_newcs} \includegraphics[height=5.5cm]{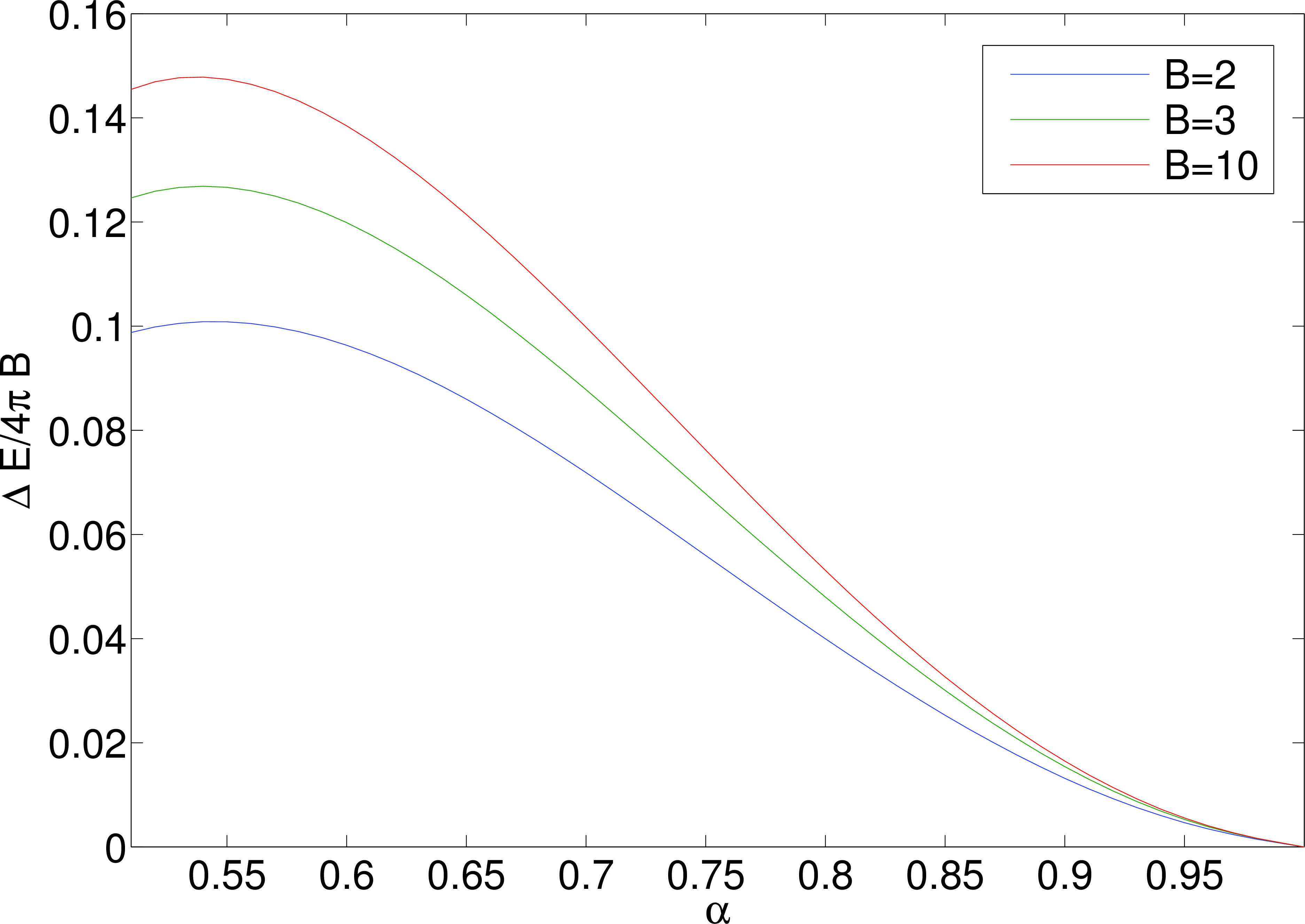}  }
        \caption{Energy as a function of model parameter $\alpha$ for axially symmetric configurations with topological charges $B=1-3,~10$.  (a)~Total energy $E$. (b)~Binding energy per soliton $\Delta E/4\pi B$.}\label{fig:energy_overview}
\end{figure}

Figure~\ref{fig:energy_comparison_newcs} displays the total energy $E$ as a function of model parameter $\alpha$ for axially symmetric baby Skyrmions with topological charges $B=1-3,$ $10$. The topological energy bound is indicated by a dashed line. All energy values are given in units of $4\pi B$. The energy values are furthest from the bound towards $\alpha=0.5$, but draw closer as $\alpha$ increases until the bound is finally saturated in the $\alpha=1$ baby Skyrme model. As the charge increases, the difference between the energy value and the bound grows smaller. So the bound tightens for higher-charge solutions of a given model. Note that the energy difference between subsequent charges also decreases drastically. As a limiting case, we include in Fig. \ref{fig:energy_overview} the energy values for charge 10. In Fig.~\ref{fig:binding_newcs} we display the binding energy as a function of $\alpha$, where the binding energy per soliton is given by
\begin{align}\label{binding}
\frac{\Delta E}{B}=E_1-\frac{E_B}{B},
\end{align}
with $E_1$ denoting the energy of the charge one solution and $E_B$ denoting the energy of the charge $B$ solution. The binding energy per soliton is the energy required to split a charge $B$ baby Skyrmion into $B$ charge one Skyrmions divided by the total number of solitons. The binding energy is found to increase with the topological charge.

\begin{figure}[t]
\begin{center}
\begin{includegraphics}[width=10cm]{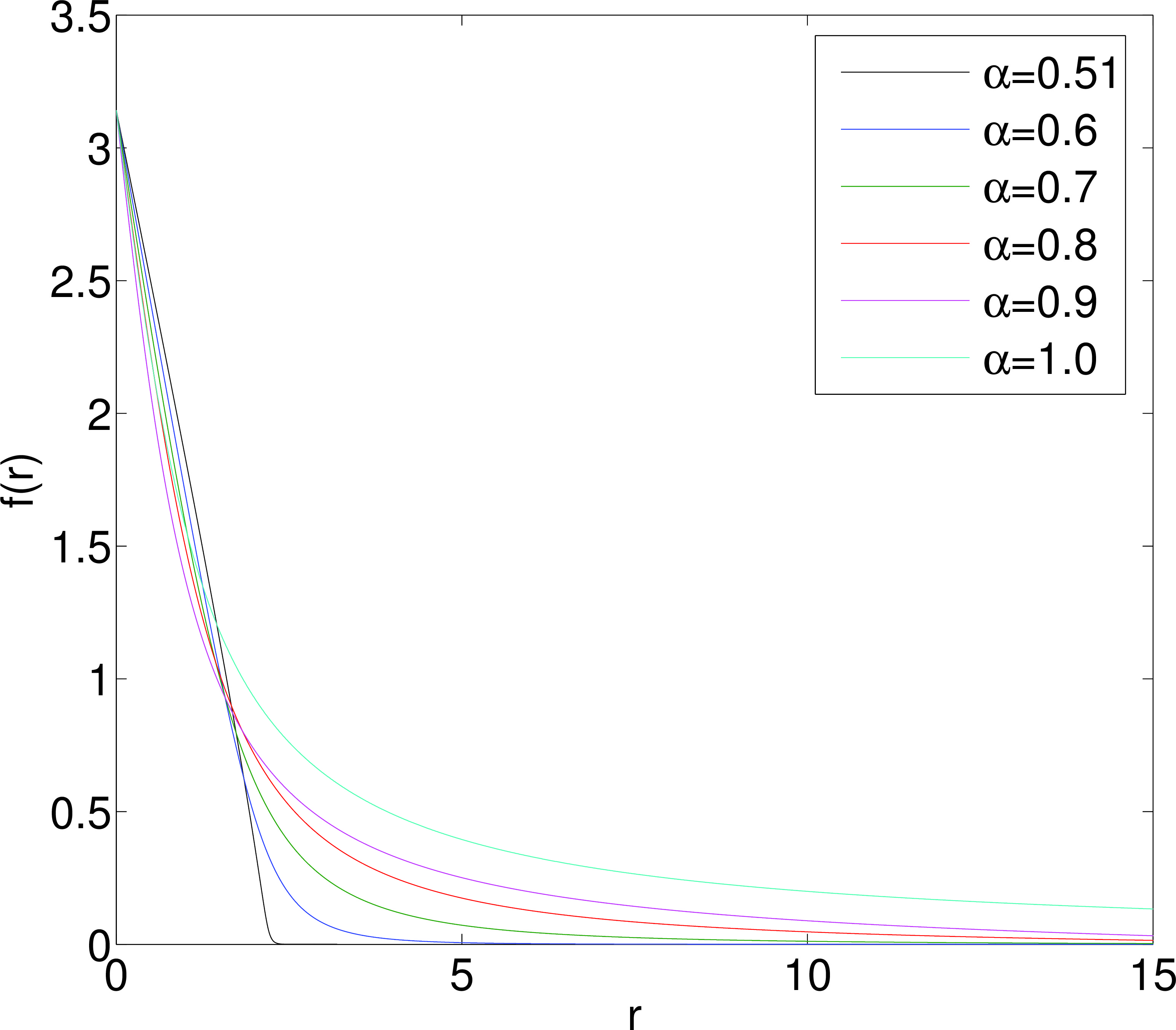}
\end{includegraphics}
\end{center}
\caption{$B=1$ profile functions $f(r)$ for model parameter $0.5 < \alpha \leq 1$.}
\label{fig:profs}
\end{figure}

Another feature of the solutions which changes dramatically as $\alpha$ increases is illustrated in Fig.~\ref{fig:profs}. Here we compare charge one profile functions $f(r)$ in a selection of the models ranging from $\alpha=0.51$ to $\alpha=0.9$ with the exact solution of the Bogomolny equation for $\alpha=1$. The numerical profile functions were calculated using the Newton method over the interval $0\leq r \leq 20$. For $\alpha=0.51$, the profile function is tightly concentrated between $r=0$ and $r=2.2$. As $\alpha$ increases, the profile functions start to spread out. By $\alpha=1$, the profile function is less localized and approaches the vacuum gradually.

A more detailed examination of the approach to the vacuum of the profile functions is given in Appendix~\ref{app:near_f0}. In this appendix, we linearize the equation of motion as $r\rightarrow\infty$ and obtain solutions that describe the profile functions as they approach zero. The profile functions exhibit a power law behavior, $f(r)\sim r^\lambda$ for large $r$. As $\alpha$ tends to $0.5$ the exponent $\lambda$ becomes increasingly negative, and the approach to the vacuum becomes steeper.  At $\alpha=0.5$ the exponent diverges suggesting that solutions in this model are compactons. Near the origin, for all values of $\alpha$ the charge one profile functions exhibit linear behavior. This is discussed in detail in Appendix~\ref{app:near_r0} where we linearize the equation of motion near the origin for any charge $B$.

%---------------------------------------------------------------------------------------

\subsection{Baby Skyrme solutions in the $\alpha=0.5$ model}\label{sec:half_sols}
 In this section, we present the results of 2D simulations for the compact charge one and two solitons obtained for model parameter $\alpha=0.5$. As a starting point for our 2D energy minimization routine we choose two different initial conditions: a rotationally symmetric configuration created from the 1D profile function for $\alpha=0.5$ and the configuration relaxed with $\alpha=0.51$.

To find profile functions in the $\alpha=0.5$ model we minimize the energy \eqref{ansatz-energy} over intervals $r$ $\in$ $[0,R_{\text{est}}]$ for various boundary points $R_{\text{est}}$ surrounding the expected compacton radius. Solving the corresponding field equation over each interval is accomplished by Newton's method for nonlinear systems with grid spacing $\Delta r=10^{-2}$ due to its increased speed over the gradient flow method. We then seek the value of $R_{\text{est}}$ that minimizes the energy. This enables us to decide upon a numerical energy value for the solution up to one decimal place of accuracy, with the virial theorem and topological charge also correct to one decimal place.

By substituting the profile functions obtained by this method into the axially symmetric ansatz \eqref{ansatz}, we build 2D configurations. These are implemented as initial configurations in a 2D relaxation method similar to that described in Ref.~\cite{Battye:2013tka}. We evolve the equations of motion derived from \eqref{final_en} in a fictitious time $t$ and include a damping term governed by the dissipation $\epsilon$. We periodically remove kinetic energy by setting $\dot{\bphi}=0$ at all grid points. In the following, the finite difference approximations are second-order accurate in the spatial derivatives. The simulations are performed on a $(401)^2$ grid with spacing $\Delta x$$=$$0.02$ for the charge one soliton and  $\Delta x=0.04$ for the charge two soliton. In both cases the dissipation parameter $\epsilon$ is set to 0.5.

\begin{figure}
        \begin{center}
     \subfigure[]{\includegraphics[height=6cm]{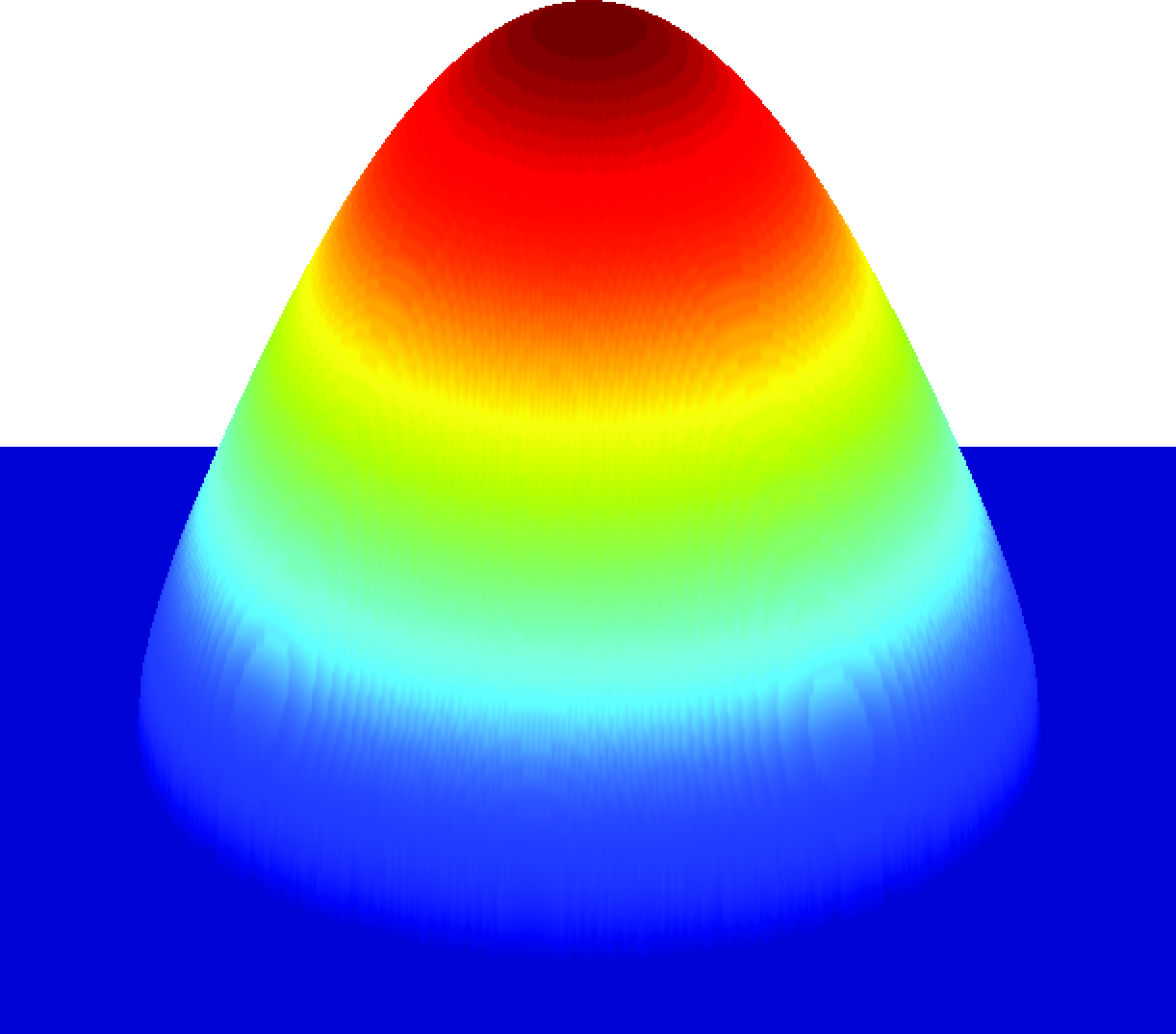}\label{fig:a12}}
       ~~   \subfigure[]{\label{fig:slices}\includegraphics[height=6cm]{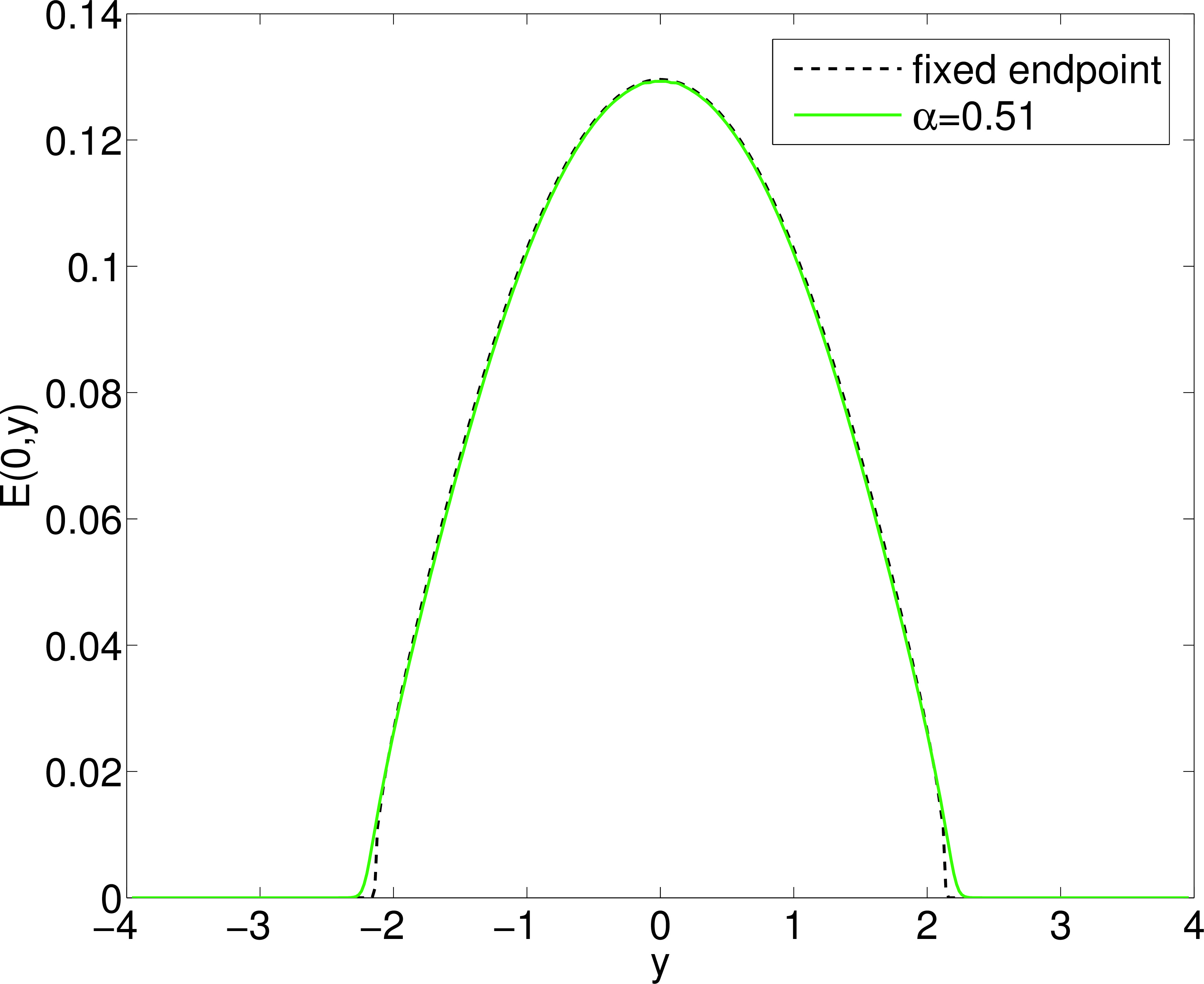}}
        \end{center}
        \caption{Energy density for charge one baby Skyrmions with model parameter $\alpha=0.5$. (a)~Surface plot of the energy density. (b)~We compare slices through the energy density obtained when relaxing two different $B=1$ initial conditions: a baby Skyrme configuration relaxed with $\alpha=0.51$ (green line) and a rotationally symmetric configuration generated from an $\alpha=0.5$ profile function.}\label{fig:a12_overview}
\end{figure}

\begin{figure}
        \begin{center}
     \subfigure[]{\includegraphics[height=6cm]{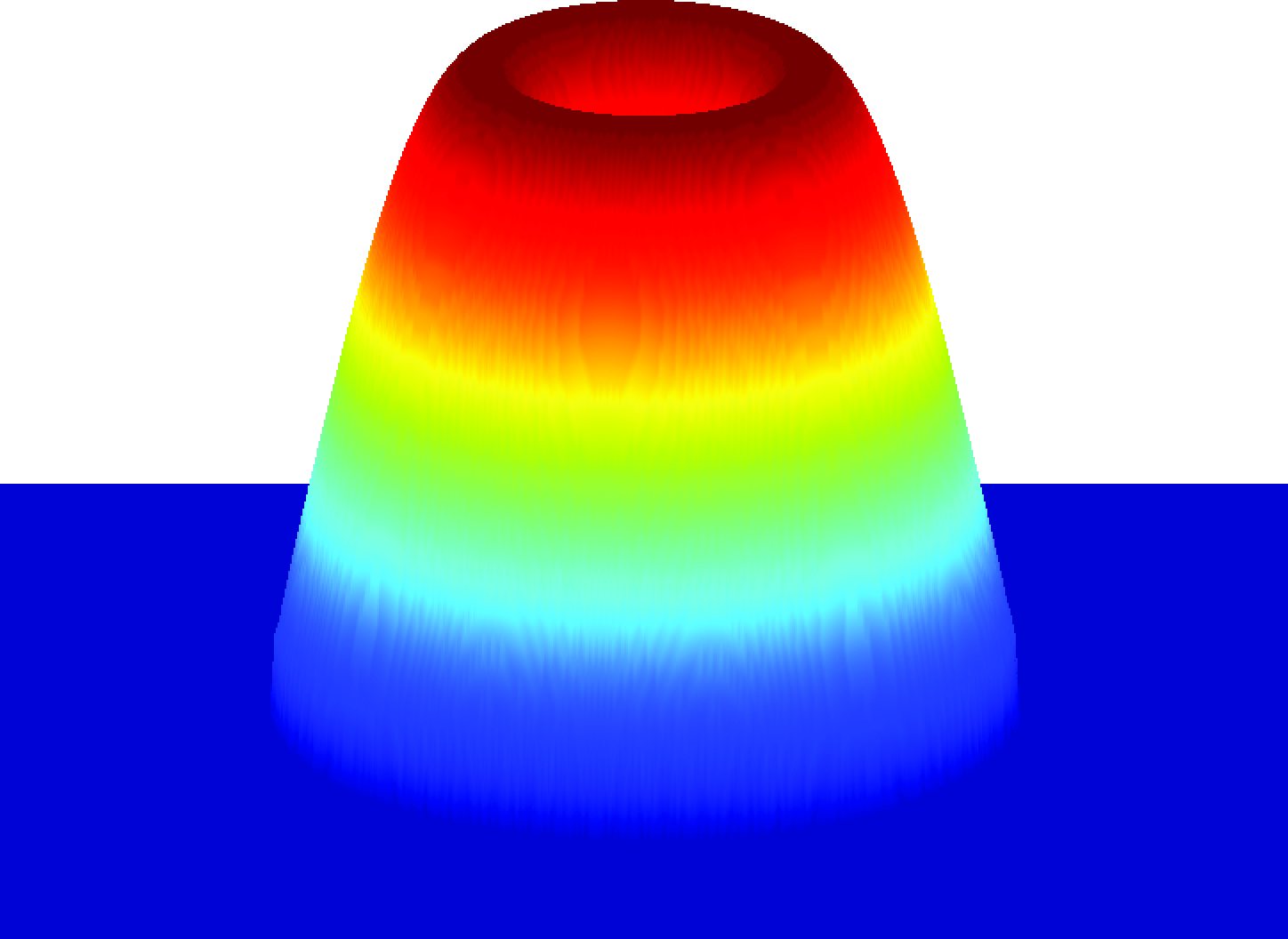}\label{fig:a12B2}}
       ~~   \subfigure[]{\label{fig:slicesB2}\includegraphics[height=6cm]{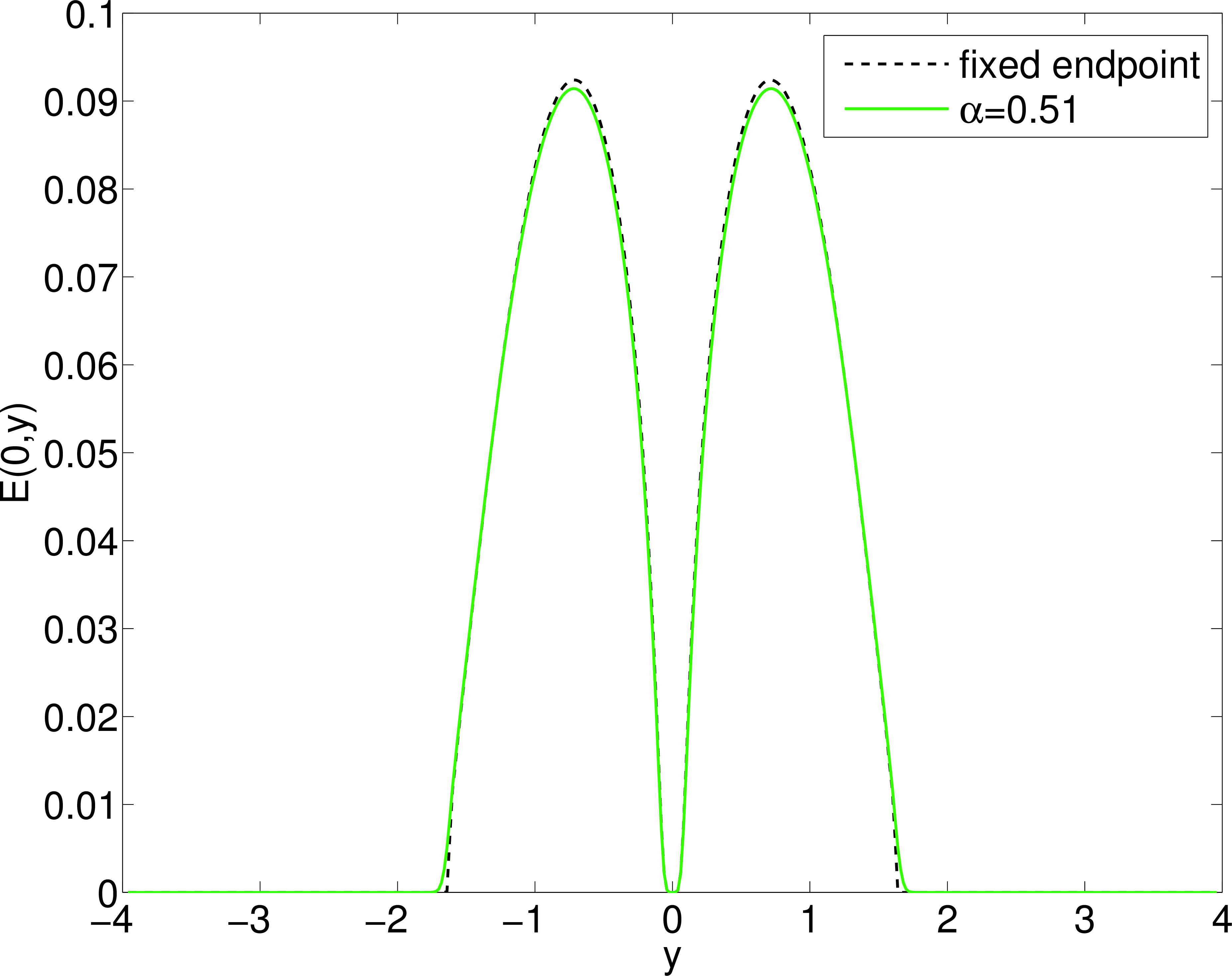}}
        \end{center}
        \caption{Energy density for charge two baby Skyrmions with model parameter $\alpha=0.5$. (a)~Surface plot of the energy density. (b)~We compare slices through the energy density obtained when relaxing two different $B=2$ initial conditions: a baby Skyrme configuration relaxed with $\alpha=0.51$ (green line) and a rotationally symmetric configuration generated from an $\alpha=0.5$ profile function.}\label{fig:a12_B2_overview}
\end{figure}

In Fig. \ref{fig:a12} we plot the energy density of the resulting $B=1$ configuration. Figure \ref{fig:a12B2} displays the energy density of a $B=2$ configuration obtained by the same method. The energy of both solutions is localized in a finite region of space, and the steep approach to the vacuum is evident at the boundaries of the compactons. The energy values for these solutions agree with those of the corresponding profile functions to one decimal place and are given in Table \ref{table:energy}. 

%--- Method 2 ----%

The second method that we implement to find solutions in this model is to take a 2D configuration with $\alpha=0.51$ as an initial condition in the 2D relaxation code. The same grid and spacing are chosen as for the previous initial configuration.  In Fig. \ref{fig:slices} we compare charge one solutions obtained by both methods. We display slices along $x=0$ through their energy density. The same comparison for charge two solitons is presented in Fig. \ref{fig:slicesB2}. While the approach to the vacuum is not as steep for the second method, both methods generally agree well and describe the soliton's energy to one decimal place.

%---------------------------------------------------------------------------------------

\subsection{Higher charge solutions}\label{sec:higher_charge}

 To verify our axially symmetric charge one and two solutions and to investigate solutions of higher charge, we implement a 2D numerical method. We apply the same relaxation method as in the previous section, but with a different grid. For models excluding $\alpha=0.5$, we use a $(201)^2$ grid with spacing $\Delta x = 0.2$. To create initial configurations, we substitute our numerical profile functions into the axial ansatz \eqref{ansatz} to generate 2D configurations. We then take these configurations as initial conditions for our 2D energy minimization algorithm to find solitons of different $\alpha$ values. For example, an $\alpha=0.8$ axial solution is  chosen as an initial configuration to obtain the $\alpha=0.7$ solution.

\bgroup
\def\arraystretch{1.22}
\begin{table}[ht]
\caption{Energy values obtained from full field simulations ($E^{(2D)}/4\pi B$) when compared to 1D gradient flow results ($E^{(1D)}/4\pi B$). The binding energy per soliton $\Delta E/4 \pi B$ is calculated by \eqref{binding} using the numerical 2D energy results. The two $B=3$ configurations given are the axial solution \eqref{ansatz} and the chain configuration of Fig. \ref{fig:a35-910chain} which is denoted by 3*.}
\centering
\begin{tabular}{c c c c c c c}
 & & & & & \\
\hline
$\alpha$ & $\beta$  & $B$ & $E^{(1D)}/4\pi B$ &  $E^{(2D)}/4\pi B$ & $\Delta E/4\pi B$   \\ [0.5ex]
\hline\hline
0.5 & 0.75 & 1 &  1.2  & 1.2 & 0.0 \\ [0.5ex]
0.5 & 0.75 & 2 &  1.1  & 1.1 & 0.1 \\ [0.5ex]
\hline
0.6 & 0.7 & 1 & 1.188  & 1.188 & 0.0 \\  [0.5ex]
0.6 & 0.7 & 2 & 1.092  & 1.092 & 0.096 \\ [0.5ex]
0.6 & 0.7 & 3 & 1.068  & 1.069 & 0.119 \\ [0.5ex]
0.6 & 0.7 & 3* & --- & 1.081 & 0.107  \\ [0.5ex]
\hline
0.7 & 0.65 & 1  & 1.130  & 1.130 & 0.0 \\ [0.5ex]
0.7 & 0.65 & 2  & 1.059   &  1.058 & 0.072 \\ [0.5ex]
0.7 & 0.65 & 3 & 1.043  & 1.042 & 0.088  \\ [0.5ex]
0.7 & 0.65 & 3* & ---  &   1.056 & 0.074 \\ [0.5ex]
\hline
0.8 & 0.6 & 1  & 1.068  & 1.068 &  0.0 \\ [0.5ex]
0.8 & 0.6 & 2  & 1.028  & 1.029 & 0.039 \\ [0.5ex]
0.8 & 0.6 & 3 & 1.020  & 1.020 & 0.048 \\ [0.5ex]
0.8 & 0.6 & 3* & --- & 1.036  & 0.032 \\ [0.5ex]
\hline
0.9 & 0.55 & 1  & 1.021  & 1.020 & 0.0 \\ [0.5ex]
0.9 & 0.55 & 2  & 1.008  & 1.007 & 0.013 \\ [0.5ex]
0.9 & 0.55 & 3 & 1.005  & 1.005 & 0.015  \\ [1ex]
0.9 & 0.55 & 3* & --- & 1.009 & 0.011  \\ [0.5ex]
\hline
\end{tabular}
\label{table:energy}
\end{table}

In Table \ref{table:energy} we give the energy values of our numerical simulations for a selection of $\alpha$ values. All energy values are given in units of $4\pi B$, motivated by the energy bound of our models \eqref{4pi_bound}. We also present the binding energy for the 2D configurations, calculated using \eqref{binding}. For axial solutions the results of our 2D simulations agree to between two and three decimal places with the values obtained when minimizing \eqref{ansatz-energy}. 

For higher charges, axial solutions remain the energetic minima, though other configurations have been obtained. In particular, we find chain configurations in our models by using three solitons in a line as an initial configuration. The energy values for the chain configurations are also presented in Table \ref{table:energy} and are denoted by a $*$. Their energy is higher than that of the axial configurations, and they do not satisfy the virial theorem. So these are local minima but not the global energy minimizers. 

\begin{figure}
\begin{includegraphics}[width=15cm]{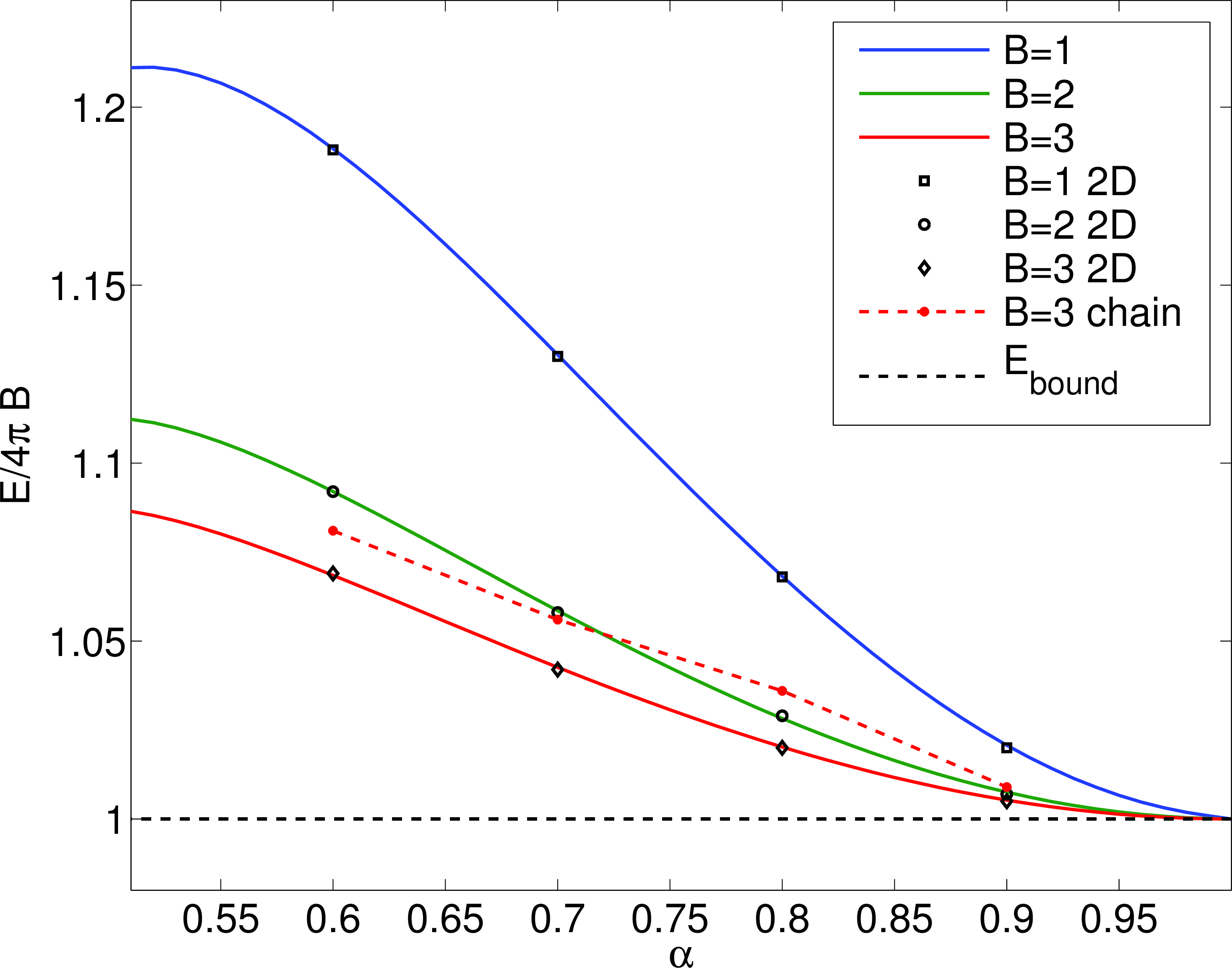}\end{includegraphics}
\caption{Total energy $E$ for baby Skyrmions obtained with the axial ansatz \eqref{ansatz} and for baby Skyrmions obtained with the 2D relaxation method.}\label{fig:en_1D2Dcomparison}
\end{figure}

In Fig. \ref{fig:en_1D2Dcomparison} we compare the energies obtained by 1D gradient flow with those calculated by 2D relaxation for baby Skyrmions with topological charges $B=1$~-~$3$. We plot the energy for axially symmetric configurations with model parameter $\alpha=0.51$~-~$1.0$ and indicate all energy values computed by the full field simulations by points.  As before, the topological energy bound is indicated by a black dashed line. The black points denote the energy of axially symmetric baby Skyrmions calculated by full field relaxation. They lie on top of the lines showing the energy for baby Skyrmions of the same topological charge obtained by 1D gradient flow. The energy for charge three chain configurations is also included in the figure as a series of red points connected by a dashed red line. For $\alpha=0.8$ the energy of the $B=3$ chain is slightly higher than the $B=2$ solution, but much lower than the $B=1$ solution. So, it is \emph{not} energetically favorable for the chain to split up into a $B=1$ and $B=2$ Skyrmion. 

The energy density for $B=1$~-~$3$ baby Skyrmions with $\alpha=0.6$~-~$0.9$ is plotted in Fig.~\ref{fig:a12-910charge1-3}. They are all axially symmetric, and the effect of increasing $\alpha$ on the solutions can be seen by comparing the graphs. This is most noticeable for the charge one solitons, where the energy density of the $\alpha=0.6$ solution is concentrated over a wide area with only a small tail. As $\alpha$ increases, the tail of the energy density becomes wider while the area in which the energy density is most concentrated decreases in width and increases in height. A similar effect occurs for the charge two and three rings, which become thinner and taller as $\alpha$ increases.

 In Fig.~\ref{fig:a35-910chain}, we display the energy density of chain configurations for the same selection of models. The structure of the chain configurations changes significantly as $\alpha$ increases. For $\alpha=0.6$, the solitons are very close together but as $\alpha$ increases, the chain starts to pull apart. For $\alpha=0.9$, the chain almost splits into three separate solitons, though they remain close enough to deform each other. This may be explained by the approach to the Bogomolny solutions at $\alpha=1$. The attraction between solitons becomes weaker and weaker until they do not feel any attraction or repulsion at $\alpha=1$. Then the baby Skyrmions can be placed at arbitrary positions. In fact the solution space is the space of based rational maps. In the $\alpha=1$ model, the energy of three separate solitons is identical to that of a three-soliton ring solution.  

 The chains observed at the lower $\alpha$ values most resemble those found in the baby Skyrme model \cite{Foster:2009vk}, although the $\alpha=0.6$ chain in particular appears more squashed. The chains observed at higher $\alpha$ values are quite different. However, comparisons could be drawn between the $\alpha=0.9$ chain and the isospinning baby Skyrmions of \cite{Battye:2013tka} in which chains are also seen to break up.

\clearpage

\begin{figure}[h]
\begin{center}
\begin{includegraphics}[width=17cm]{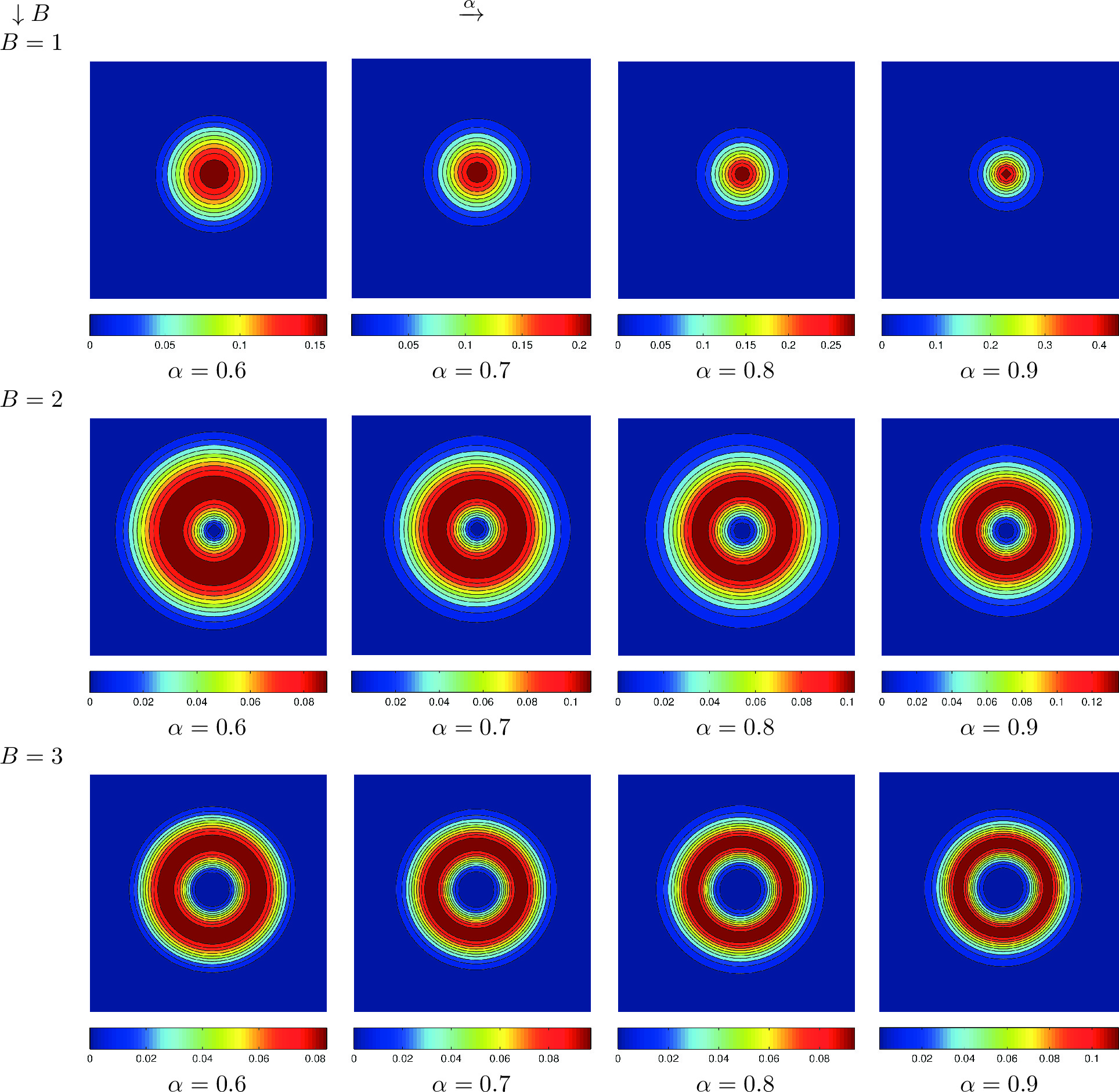}\end{includegraphics}
\end{center}
\caption{Energy density contour plots for baby Skyrmions with model parameter $\alpha~=~0.6,~0.7,~0.8,~0.9$ and charges $B=1-3$. }
\label{fig:a12-910charge1-3}
\end{figure}

\begin{figure}[h]
\begin{center}
\begin{includegraphics}[height=18cm]{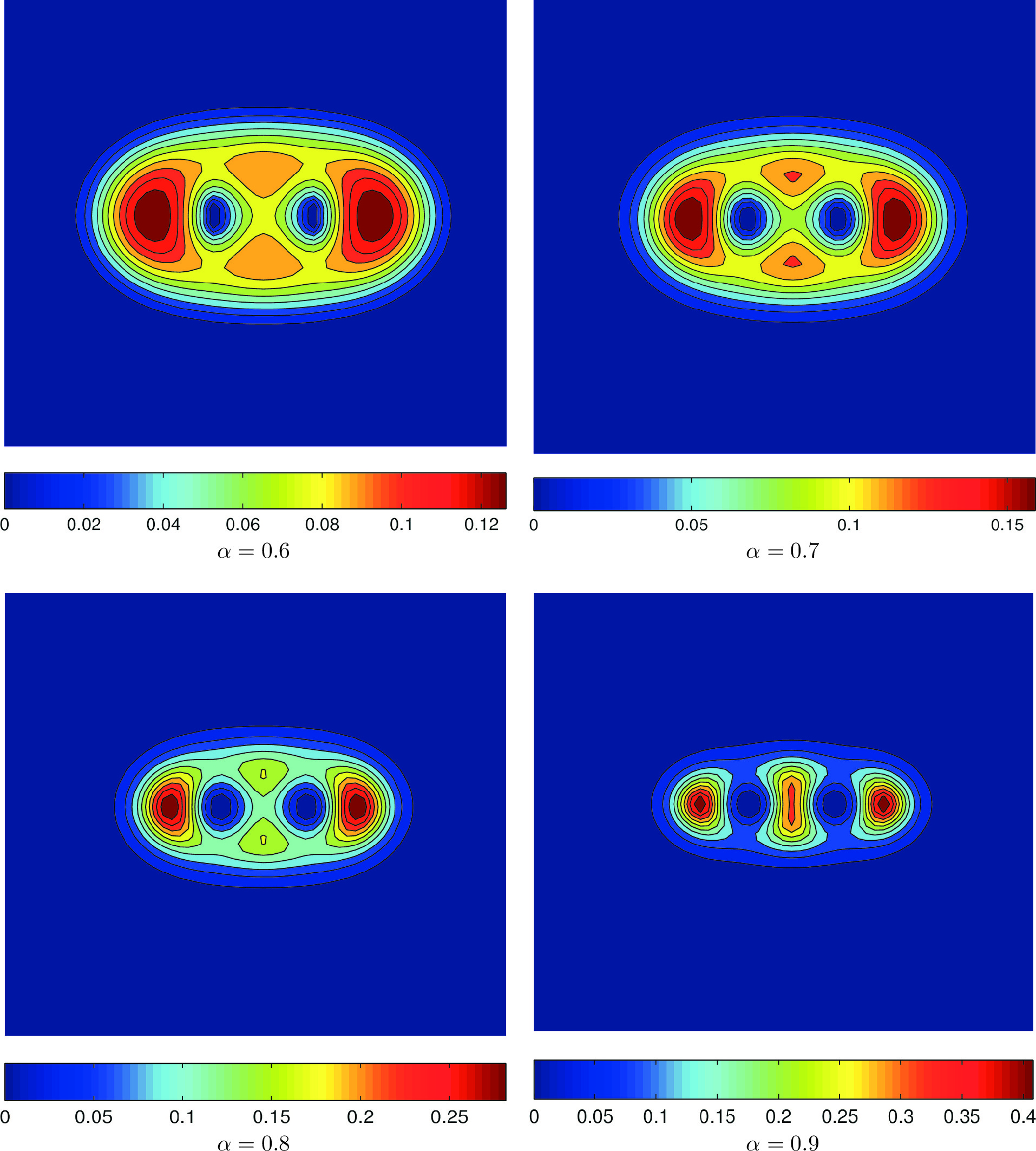}\end{includegraphics}
\caption{Contour plots of the energy density for $B=3$ chain configurations. The value of $\alpha$ for each configuration is indicated underneath its plot.}
\label{fig:a35-910chain}
\end{center}
\end{figure}

\clearpage

%%%%%%%%%%%%%%%%%%%%%%%%%%%%%%%%%%%%%%%%%%%%%%%%%%%%%%%%%%%%%%%%%%%%%%%%%%%%%%%%%%%%%%%%%

\section{Conclusions}\label{sec:conc}

We have developed a one-parameter family of baby Skyrme models that do not
require a potential term to admit topological solitons. Starting with a
general form for our models involving four parameters, we fixed three of
these by specifying that our models should satisfy the topological energy
bound $E\geq 4\pi|B|$.  Similarly to the Skyrme models described in
\cite{Gillard:2015eia}, we found that at one end of our parameter range
there is a model in which this bound can be saturated. This model is scale
invariant, and exact solutions to the Bogomolny equations can be obtained
for any topological charge. Furthermore, our choice of parameters ensures that all of our models satisfy the same virial theorem, $E_2=E_4$, as the Skyrme model. In this way, we have designed a one-parameter family of baby Skyrme models without a potential whose scaling behavior better matches the
Skyrme model and which even includes a baby Skyrme model scaling exactly
like the Skyrme model at the extreme of our parameter range where
$\alpha=0.5$.

Our investigation into the solitons of our models showed that their form
greatly depends upon the choice of the parameter $\alpha$. Solitons in the
$\alpha=0.5$ model are compactons. Both our numerical results for this
model and linearization arguments support this conclusion. Previous
examples of compactons in baby Skyrme models
\cite{Adam:2009px,Adam:2010jr,Speight:2010sy} depend on the choice of
potential term used, and typically occur for particular parameter values
in a one-parameter family of potential functions. There is no potential
term in our models but the importance of parameter choice to the existence
of compactons is similarly observed here.

Solitons in the models with $\alpha > 0.5$ were found to be less
localized. We calculated solutions numerically for a selection of the
models using three different methods. The energy of
solutions to our models decreases as $\alpha$ increases, and higher-charge solutions are more tightly bound near the $\alpha=0.5$ end of the
parameter range. As we approach the extreme of the parameter range in
which the energy bound can be saturated, the binding energy of solutions
decreases to zero. The energy minimizers are
axially symmetric solutions, even for topological charge three.
However, we also observed other higher-charge configurations with greater
energy, in particular $B=3$ chain configurations. Chain solutions for
$\alpha\in[0.6,0.7]$ most closely resemble those observed in the old baby
Skyrme model \cite{Foster:2009vk}, while other chains have a very
different appearance. In particular, as $\alpha$ increases our chain
configurations begin to pull apart and become three almost separate
solitons.

In this paper we have provided an initial study of our new baby Skyrme
models but there is still further work that could be done. One interesting
question is how
the solutions behave for higher charges. Will the axially symmetric
solutions always be the minimum energy configurations or will chain-like
configurations play a more prominent role? An important challenge is to
develop more accurate numerical methods to calculate compacton solutions,
as this may have important applications beyond the area of topological
solitons, for example in relation to fractional Laplacians \cite{Huang}
and fractional diffusion \cite{Ervin}.
Although the unique aspect of our models is their lack of need for a
potential term, the inclusion of a potential term in this setting offers
many possibilities. Following \cite{Adam:2013tga} and balancing the different terms with
the potential would give rise to multi-parameter families of models which
still obey a linear energy bound.  The choice of potential has an
important effect on the structure of solitons in the usual baby Skyrme
model. Whether the same is true of our new models and generalizations
thereof remains to be determined.

%%%%%%%%%%%%%%%%%%%%%%%%%%%%%%%%%%%%%%%%%%%%%%%%%%%%%%%%%%%%%%%%%%%%%%%%%%%%%%%%%%%%%%%%%
%%%%%%%%%%%%%%%%%%%%%%%%%%%%%%%%%%%%%%%%%%%%%%%%%%%%%%%%%%%%%%%%%%%%%%%%%%%%%%%%%%%%%%%%%

\section*{Acknowledgements}
The authors would like to thank David Foster for sparking their interest in constructing baby Skyrmions without potential terms and for fruitful discussions at various stages of this project. We are also grateful to Andrzej Wereszczynski for pointing out the generalized version of the topological energy bound \eqref{generalbound} and to other participants of the conference ``Solitons: Topology, Geometry and Applications" at Thessaloniki in April 2015 for useful comments. J.~E.~A. acknowledges the UK Engineering and Physical Science Research Council (Doctoral Training Grant Ref. EP/K50306X/1) and the University of Kent School of Mathematics, Statistics and Actuarial Science for a Ph.D studentship.  The work of S.~K. and M.~H. was financially supported by the UK Engineering and Physical Science Research Council (Grant No. EP/I034491/1). M.~H. was partly funded by the UK Science and Technology Facilities Council under Grant No. ST/J000434/1.

\appendix

\section{Linearizing as $r\rightarrow\infty$}\label{app:near_f0}
In this appendix, we linearize the equation of motion for an  axially symmetric charge $B$ soliton in the baby Skyrme model \eqref{energy-alphabeta} for large $r$. As discussed in Secs. \ref{sec:Derrick} and \ref{sec:bounds}, we set $\beta=1-\frac{\alpha}{2}$, and the constants $c_1$ and $c_2$ are fixed in \eqref{c1c2}.

The equation of motion for the profile function $f(r)$ is given by
\begin{align}
\label{symm-eqm-alphabeta}
f'' & \Bigg(\frac{4\alpha(\alpha-1)c_1f'^2}{f'^2+\frac{B^2}{r^2}\sin^2f} +\frac{2\beta(2\beta-1)c_2\big(2f'^2\frac{B^2}{r^2}\sin^2f\big)^\beta\big(f'^2+\frac{B^2}{r^2}\sin^2f\big)^{1-\alpha}}{f'^2} +2\alpha c_1 \Bigg) \nonumber \\
+ & f'\Bigg(\frac{2\alpha c_1}{r} + \frac{2\alpha(\alpha-1)c_1\big(-2\frac{B^2}{r^3}\sin^2f + \frac{B^2}{r^2}f'\sin 2f\big)}{f'^2+\frac{B^2}{r^2}\sin^2f}\Bigg) \nonumber \\
+ & \frac{c_2\beta(2\beta-1)\big(2f'^2\frac{B^2}{r^2}\sin^2f\big)^\beta\big(f'^2+\frac{B^2}{r^2}\sin^2f\big)^{1-\alpha}\big(-\frac{2}{r}\sin^2f + f'\sin 2f\big)}{f'\sin^2f}  \nonumber \\
- & \alpha c_1\frac{B^2}{r^2}\sin 2f  =0.
\end{align}

\begin{figure}[t]
\begin{center}
\begin{includegraphics}[width=8cm]{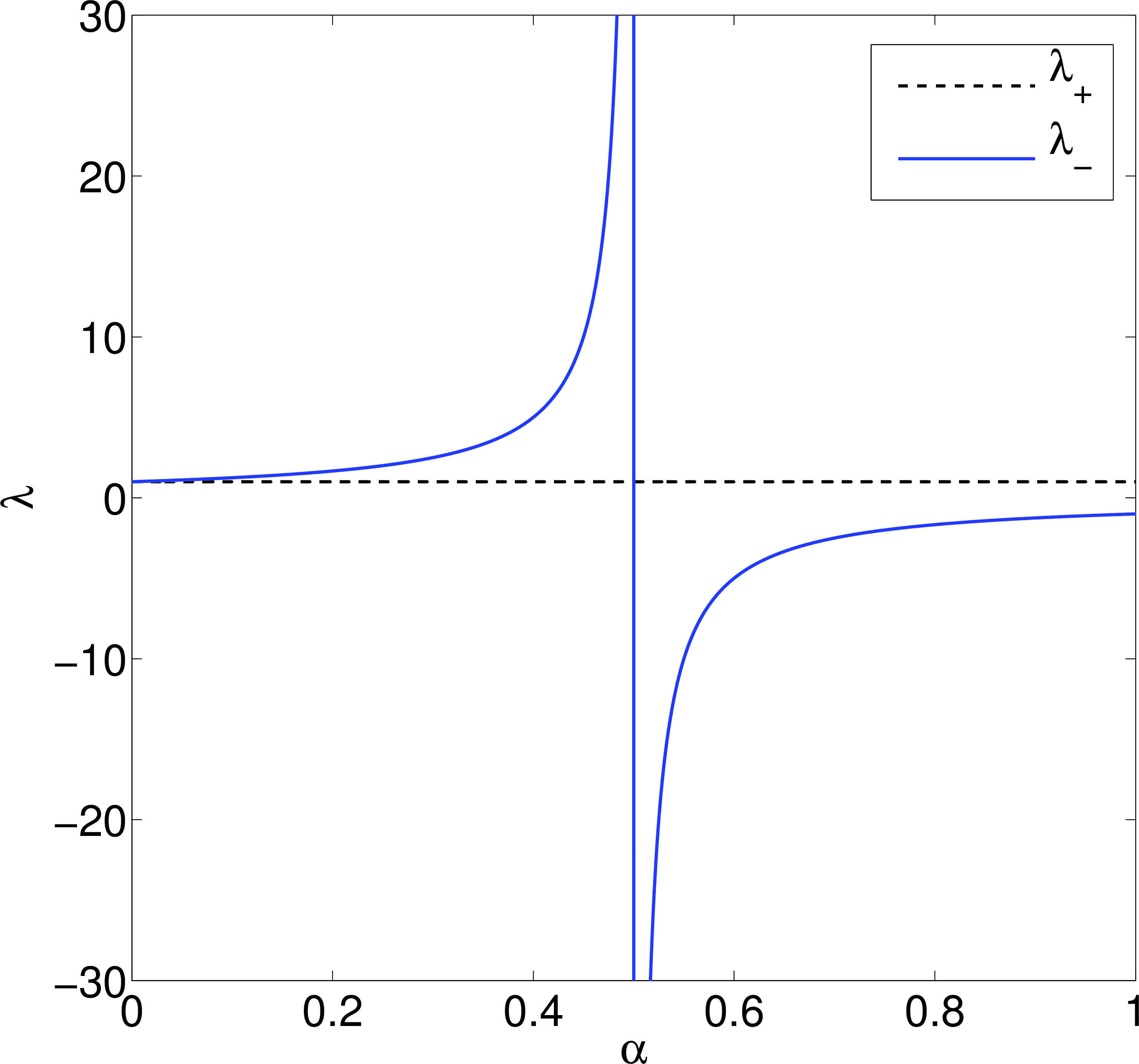}
\end{includegraphics}
\caption{Real solutions $\lambda_+,~\lambda_-$ of Eq. \eqref{cancel_lin} as functions of $\alpha$ for $0<\alpha<1$ and $B=1$.}
\label{fig:lin}
\end{center}
\end{figure}

To linearize the symmetric equation of motion \eqref{symm-eqm-alphabeta} as $r\rightarrow \infty$, we substitute $f(r)=r^\lambda$ into \eqref{symm-eqm-alphabeta} and consider only the leading-order terms. For $\alpha\leq1$, this leads to the equation
\begin{align}\label{cancel_lin}
\frac{(2\alpha-1)\lambda^4-2(\alpha-1)\lambda^3+2B^2(\alpha-1)\lambda^2-2B^2(\alpha-1)\lambda-B^4}{\lambda^2+B^2}=0,
\end{align}
which simplifies to the quadratic
\begin{align}\label{cancel_lin2}
(2\alpha-1)\lambda^2-2(\alpha-1)\lambda-B^2  =  0 .
\end{align}
For $\alpha\neq\frac{1}{2}$ this can be solved for $\lambda$, to find
\begin{align}\label{lroots}
\lambda_\pm = \frac{\alpha-1\pm\sqrt{\alpha^2+2(B^2-1)\alpha-B^2+1}}{2\alpha-1}.
\end{align}
For parameter value $\alpha=1$, most of the terms cancel and the solutions are
\begin{align}\label{a1_linssol}
\lambda=\pm B.
\end{align}
 At the other extreme, where $\alpha=0.5$, the quadratic term in \eqref{cancel_lin2} cancels. The solution $\lambda_-$ becomes singular and $\lambda_+=B^2$. In Fig. \ref{fig:lin} we plot the two solutions $\lambda_+,~\lambda_-$ of Eq. \eqref{cancel_lin2} as a function of the model parameter $\alpha$ for $0<\alpha<1$ and topological charge $B=1$. The root $\lambda_+$ is always positive and therefore we discard it. The root $\lambda_-$ has an asymptote at $\alpha=0.5$ where Eq. \eqref{cancel_lin} is singular. For $0<\alpha<0.5$, the values of $\lambda_-$ are positive and therefore solutions for this range of $\alpha$ are either compact or do not have finite energy. The interesting parameter range is $0.5<\alpha\leq1$ in which $\lambda_-\leq-1$ and solutions have finite energy. 

For $\alpha>1$, the leading-order terms in \eqref{symm-eqm-alphabeta} after substituting $f(r)=r^\lambda$ are different, so that a different linearized equation is found in this parameter range. Here the equation becomes
\begin{align}\label{a>1_eq}
\frac{2(2-\alpha)(1-\alpha)c_2}{\lambda}(\lambda-1)\big(2\lambda^2B^2\big)^{1-\alpha/2}
\big(\lambda^2+B^2\big)^{1-\alpha}=0
\end{align}
The only solutions are $\lambda=1$ and $\lambda=\pm iB$. Thus for  $\alpha>1$,  solutions to the equation of motion would either have infinite energy or be compact.

%%%%%%%%%%%%%%%%%%%%%%%%%%%%%%%%%%%%%%%%%%%%%%%%%%%%%%%%%%%%%%%%%%%%%%%%%%%%%%%%%%%%%%%

\section{Linearizing near $r=0$}\label{app:near_r0}
We also linearize the equation near $r=0$ to gain a greater understanding of the behavior of solutions in relation to their topological charge. Set
\begin{align}\label{lin_subs}
 f(r) = \pi- ar^\gamma,
\end{align}
in Eq. \eqref{symm-eqm-alphabeta}, where $a$ is constant, and assume that $\gamma\geq1$. For small $r^\gamma$, we can use the small angle approximation to replace trigonometric terms. By our assumption on $\gamma$ and the constraint $0.5\leq\alpha\leq1$, we find the leading-order equation
\begin{align}\label{leading_ord}
2\alpha ac_1\left(2\gamma^3(\alpha-1)\frac{1-\gamma}{B^2+\gamma^2}+\gamma(1-\gamma)-2\gamma(\alpha-1)B^2\frac{\gamma-1}{B^2+\gamma^2}+ B^2-\gamma\right)=0,
\end{align}
which simplifies to the quadratic
\begin{align}\label{leading_ord2}
B^2+2(\alpha-1)\gamma-(2\alpha-1)\gamma^2=0.
\end{align}
For $\alpha>0.5$, we can solve this to find the positive root
\begin{align}\label{gamma}
\gamma=\frac{-1+\alpha+\sqrt{\alpha^2+2(B^2-1)\alpha+1-B^2}}{2\alpha-1}.
\end{align}
Note that once again the $\alpha=0.5$ case must be considered separately. Here we obtain the leading-order equation
\begin{align}\label{leading_ord_half}
B^2-\gamma=0,
\end{align}
so we find $\gamma=B^2$ when $\alpha=0.5$. This is also the limit of \eqref{gamma} as $\alpha\rightarrow0.5$.

At the other end of our allowed range of $\alpha$ values, where $\alpha=1$, we observe that the expression \eqref{gamma} simplifies to
\begin{align}\label{gamma_a1}
\gamma=|B|.
\end{align}

Notice that for $B=1$ and any choice of $\alpha\geq0.5$, we find $\gamma=1$. So for all of our models, the charge one profile function has a linear behavior near the origin. For $B>1$  we confirm that $\gamma(\alpha)$ as given by \eqref{gamma} does not have any turning points in the interval $0.5<\alpha<1$, with $\gamma(0.5)=B^2$ and $\gamma(1)=|B|$. Hence $\gamma\geq1$ for any choice of $B\geq1$. This justifies our earlier assumption on $\gamma$.

%%%%%%%%%%%%%%%%%%%%%%%%%%%%%%%%%%%%%%%%%%%%%%%%%%%%%%%%%%%%%%%%%%%%%%%%%%%%%%%%%%%%%%%%%
%%%%%%%%%%%%%%%%%%%%%%%%%%%%%%%%%%%%%%%%%%%%%%%%%%%%%%%%%%%%%%%%%%%%%%%%%%%%%%%%%%%%%%%%%

\bibliography{mybib}

%%%%%%%%%%%%%%%%%%%%%%%%%%%%%%%%%%%%%%%%%%%%%%%%%%%%%%%%%%%%%%%%%%%%%%%%%%%%%%%%%%%%%%%%
%%%%%%%%%%%%%%%%%%%%%%%%%%%%%%%%%%%%%%%%%%%%%%%%%%%%%%%%%%%%%%%%%%%%%%%%%%%%%%%%%%%%%%%%

\end{document}